\documentclass[journal=jpccck,manuscript=article,layout=preprint]{achemso}
\setkeys{acs}{usetitle = true, doi = true, maxauthors=99}
\usepackage[T1]{fontenc}
\usepackage{newtxtext,newtxmath}
\usepackage{xcolor}
\usepackage{booktabs}
\usepackage[version=4]{mhchem}
\usepackage{multirow}
\usepackage{xr-hyper}
\usepackage{hyperref}

\makeatletter

\newcommand*{\addFileDependency}[1]{
\typeout{(#1)}
\@addtofilelist{#1}
\IfFileExists{#1}{}{\typeout{No file #1.}}
}\makeatother

\newcommand*{\myexternaldocument}[1]{%
\externaldocument{#1}%
\addFileDependency{#1.tex}%
\addFileDependency{#1.aux}%
}

\myexternaldocument{SI}

\newcommand{\angstrom}{\mbox{\normalfont\AA}}

\title{Elucidating Trace Gas Interactions with Ice Surfaces using Molecular Dynamics}
\author{Benjamin M. Harless}
\author{J. Daniel Gezelter}
\email{gezelter@nd.edu}
\affiliation[University of Notre Dame]{251 Nieuwland Science Hall, Department of Chemistry and Biochemistry, \\
University of Notre Dame, Notre Dame, Indiana 46556, USA}

\begin{document}

\begin{tocentry}
\centering\includegraphics[width=\linewidth]{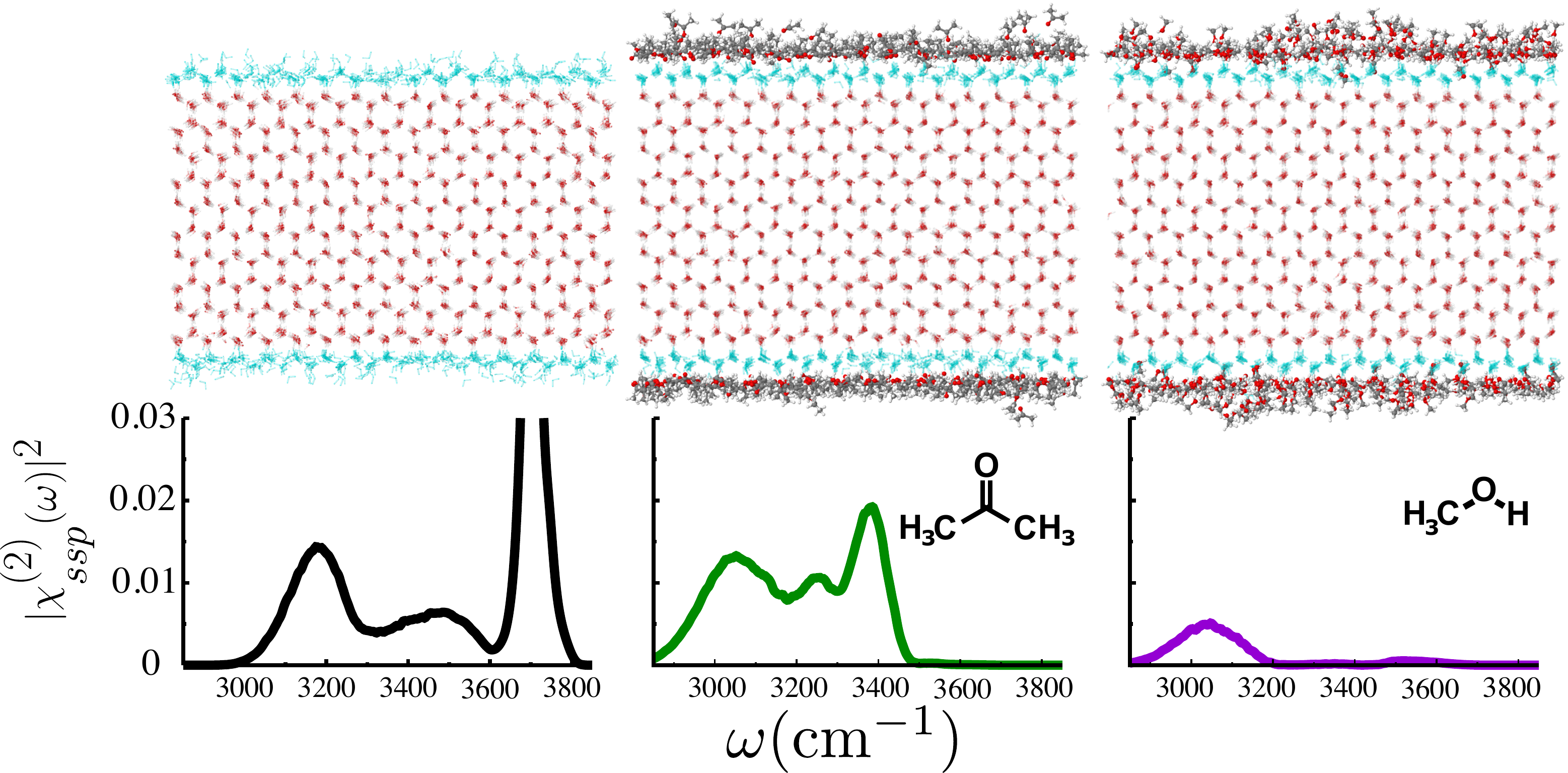} 
\end{tocentry}

\begin{abstract}
Recent experimental work carried out by Yettapu, Manning, and Cyran on ice / vapor interfaces revealed changes in sum frequency generation (SFG) signals at $\sim 3170 \mathrm{~cm}^{-1}$ depending on whether acetone or methanol was adsorbed onto the ice at 223~K. They linked the transformation of that signal to the induced melting of the ice crystal. We present molecular dynamics modeling of their experiments using methanol and acetone layers adsorbed on the basal facet of a  proton-ordered Ice-Ih crystal at temperatures of 223~K and 265~K. The width of the quasi-liquid layer (QLL), O-H bond frequencies,  calculated SFG signals,  and hydrogen bonding statistics all indicate changes due to the presence of the adsorbates. While we see no significant adsorbate-based differences in QLL formation, we find at 223~K that acetone preserves under-coordinated waters that are associated with a secondary water with a low frequency O-H bond, while methanol integrates more fully into the hydrogen bonding network, resulting in a narrower distribution of O-H bond frequencies. This change in hydrogen bonding produces a local modification of the electric field, which may contribute to the difference in the experimental SFG signals at low frequencies.
\end{abstract}

\maketitle

\section{Introduction}

The interactions of ice crystals with adsorbates have significant ramifications for environmental, atmospheric, and materials science, including the nucleation and growth of water droplets and ice crystals, leading to cloud formation and precipitation.\cite{Cantrell:2005} Important atmospheric reactions also take place between adsorbates on ice / vapor interfaces.\cite{Domine:2002,Abbatt:2003aa,BartelsRausch:2014,BartelsRausch:2025} Careful crystal growth and vapor deposition, coupled with surface-sensitive spectroscopies, can help elucidate these important interactions. At the ice / vapor interface, collective oscillations of the outermost water molecule layers are often studied via sum frequency generation (SFG) spectroscopy, yielding insights about the binding sites and surface morphologies that are presented to the adsorbed molecules. Adsorbates will interact most directly with these outermost molecules and will therefore impact the observed SFG signals. 

In a recent experimental study of acetone (\ce{(CH3)2O}) and methanol (\ce{CH3OH}) adsorption on ice surfaces, Yettapu, Manning, and Cyran observed an intriguing difference in the behavior of the SFG spectra as trace gasses flowed over the surface.\cite{Yettapu:2025aa} For a bare basal plane of ice, their SFG spectra contain signals from bonded O-H stretch vibrations at $\sim 3170$, $3250$, and $3400 \mathrm{~cm}^{-1}$. The signature of dangling O-H bonds at air / water interfaces is usually taken to be a sharp feature between $3700-3800 \mathrm{~cm}^{-1}$ which is quite close to gas-phase O-H stretch frequencies. However, there is a known anti-correlation effect of O-H frequencies in water; if the dangling O-H is at $\sim 3700  \mathrm{~cm}^{-1}$, the other O-H will be red-shifted on average compared to the bulk, i.e., $\sim 3200 \mathrm{~cm}^{-1}$,\cite{Stiopkin:2011aa}  while water molecules at the interface that have both O-H bonds participating in hydrogen-bonds appear in a broad feature from $3000-3600 \mathrm{~cm}^{-1}$.\cite{Sudera2020aa}.  Of particular interest in the new experiments are changes in the $3170 \mathrm{~cm}^{-1}$ feature, which alters appearance depending on the identity and exposure time of the adsorbate gas.

Other simulations of SFG signals have found strong contributions at $\sim 3180 \mathrm{~cm}^{-1}$ from the hydrogen bonds interconnecting ice bilayers,\cite{Buch2007aa} with later studies obtaining greater agreement for the low frequency peak from proton-striped phases of the ice surface (the Fletcher pattern) rather than random proton ordering.\cite{Buch:2008aa}  At low temperatures, striped phases present alternating rows, one with dangling hydrogens and another with lone pairs from the oxygen. These distinct environments impact collective vibrations and reorientation at the surface. Reconciling experimental and computational SFG spectra of ice interfaces remains an active area of research.

Recent calculations by Rashmi and Paesani, which coupled quantum dynamics simulations with many body potentials, provide additional insight into the SFG spectra of ice / vapor interfaces.\cite{Paesani2025aa}  Restructuring of the striped phases at low temperatures was observed and characterized in terms of the reorientation of the free O-H water molecules. This helped to build a coherent picture of the molecular motions that affect observed O-H stretch modes.\cite{Paesani2025ab}  

Attributing specific molecular contributions to the low frequency region is a complex issue with a number of competing explanations. In this paper, we focus primarily on hydrogen bonding between water and the adsorbates that could potentially impact the low frequency peak.

Yettapu, Manning, and Cyran observed that the two adsorbed trace gasses produced different effects on the low frequency feature at $\sim 3170 \mathrm{~cm}^{-1}$, which also changed depending on whether the gasses were adsorbed on ice ($\sim 223$~K) or water ($\sim 273$~K) surfaces.\cite{Yettapu:2025aa} They interpret these changes as alterations in the surface hydrogen bonding behavior that are partly due to gas-induced melting of the surface. In particular, the surface of ice with adsorbed methanol produces an SFG signal that closely resembles the water-methanol surface with the low frequency peak blue-shifted by $\sim 80 \mathrm{~cm}^{-1}$, which they interpret as a potential indication of methanol-induced surface melting. In contrast, with adsorbed acetone, the $\sim 3170 \mathrm{~cm}^{-1}$ signal persists, albeit at a lower intensity, and is slightly red-shifted by $\sim 10 \mathrm{~cm}^{-1}$.

In this work, we provide some context for the low frequency peak at the ice interface with the two different adsorbates. We modeled the ice interface at two temperatures: once at 223~K, without a well-established quasi-liquid layer (QLL) where the ice maintains proton striping at the air interface, and again at 265~K, where the ice has developed a clear QLL that behaves much more like a liquid water film. The adsorbate molecules were introduced in the gas phase above the equilibrated ice surfaces and were allowed to reach equilibrium at these temperatures. We studied the structural features of the interface, notably the local tetrahedrality and density, which allowed us to compute the widths of the liquid layers.  We also used electrostatic maps to predict the distributions of frequencies that O-H bonds would contribute to surface-active spectroscopies and computed statistical information regarding the hydrogen bonding environments of the surface layers of the ice / QLL.   We have also utilized the same electrostatic maps to calculate SFG signals for an exciton model for the surface water molecules under the time averaging approximation.

\section{Methods}
All molecular dynamics (MD) simulations were carried out using OpenMD.\cite{Drisko:2024aa}  Force field parameters for the adsorbates were mostly adapted from the Generalized Amber Force Field (GAFF2)\cite{Wang:2004aa}, while partial charges on the acetone atoms were obtained using the AM1-BCC model,\cite{Jakalian:2000aa,Jakalian:2002aa,ACPYPE} and the charges on methanol were adapted from the work of Fennell, Wymer, and Mobley,\cite{Fennell:2014aa} which provides accurate descriptions of solvation free energies for alcohols. For water, the TIP4P-Ice rigid body model was selected due to its accurate representation of the melting behavior of ice Ih.\cite{Abascal:2005aa} Simulations were carried out in simulation cells with periodic boundary conditions, using the real space damped shifted force (DSF) approach to compute electrostatic interactions with a cutoff of 12 Å and a damping parameter $\alpha = 0.18~\angstrom^{-1}$.\cite{Fennell:2006aa}  All force field parameters are provided in the Supporting Information.

\subsection{System Construction and Equilibration}
At experimental temperatures $\le 180$ K, the basal face of ice-I$_\mathrm{h}$ presents stripes of protons and lone pairs to the vapor.\cite{Groenzin2007,Buch:2008aa,Nojima:2017aa}
We have constructed our initial ice configurations from an ice-I$_\mathrm{h}$ unit cell proposed
by Hirsch and Ojam\"{a}e (Structure 6) which, when replicated, reproduces these surface features.\cite{Hirsch2004} Note that \textit{all} simulations of ice structures exhibit proton ordering on the length scale of the periodic box, but in order to reproduce proton surface striping with zero dipole crystals, we have utilized proton translational ordering on a smaller length scale than other approaches. More detailed descriptions of how the crystals were constructed are given in references \citenum{Louden:2017aa} and \citenum{Louden2013a}.
As a control system, five independent replicas of the ice crystal, presenting a basal $\{0001\}$ facet to vacuum, were prepared using different random seeds to produce distinct initial velocity distributions. 

\begin{figure}
    \centering
    \includegraphics[width=\linewidth]{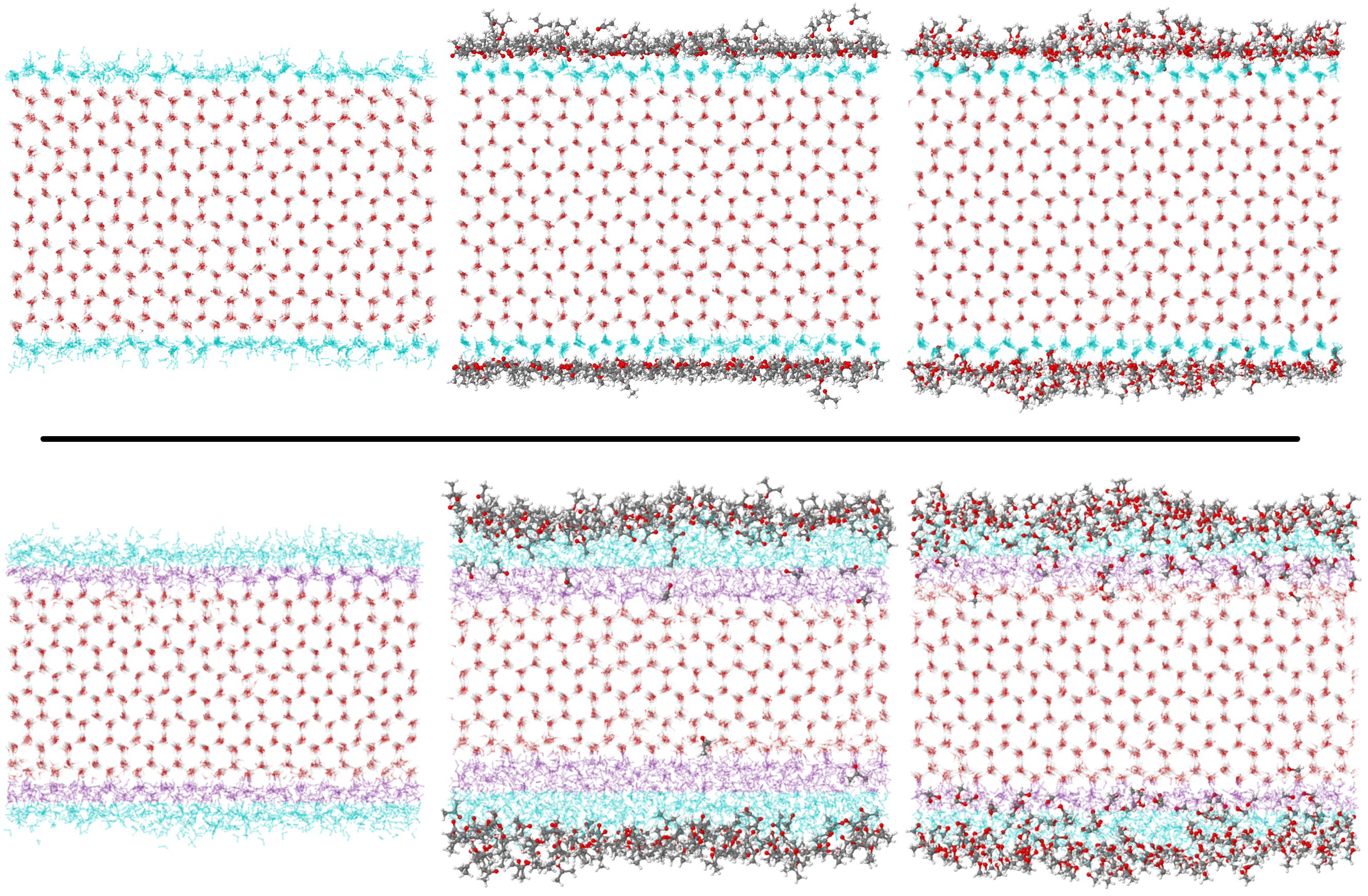} 
    \caption{The equilibrated systems at 223~K (top) and at 265~K (bottom). Bare ice (left), ice with an acetone adsorbate layer (center), ice with methanol adsorbate (right). For the 223~K surfaces, the outer bilayers of water molecules are shown in blue. For 265~K, the QLL (where $q \le 0.84$) is shown in purple and blue, although only the outer 50\% (blue) were used to sample electric fields to compute a distribution of O-H bond frequencies and to provide statistics on H-bond coordination.}
    \label{fig:systems}
\end{figure}

To assemble methanol and acetone adsorbates interacting with ice, packmol was used to generate five independent configurations for the adsorbate layers.\cite{Martinez:2009aa} To balance the total mass between the systems: 149 acetone and 270 methanol molecules were randomly placed within 10 \AA~ of either side of the basal $\{0001\}$ proton-striped interface with a large vacuum separating the adsorbate layers. Snapshots of the  low temperature (223~K) and high temperature (265~K) ice / QLL, and ice / adsorbate systems are shown in Fig. \ref{fig:systems}.

Because we were computing only structural properties of these interfaces (and not dynamical quantities), equilibration and data collection were carried out in the canonical (NVT) ensemble. Initial velocities were sampled from a Maxwell-Boltzmann distribution corresponding to 50~K. The systems underwent short structural relaxation simulations (500~fs) with 0.5 fs time steps followed by gentle heating to 265~K with a timestep of 1~fs and periodic resetting of thermostat variables (every ps) to impose gradual heating over 500~ps that kept the systems below the melting temperature of TIP4P-Ice (272 K). In order to match the temperature of the Yettapu \textit{et al.} experiments, some of the equilibration simulations ended when the temperature reached 223~K, while others proceeded to a final temperature of 265~K. After all systems had reached their target temperatures, thermostat variables were reset, and the simulations were allowed to equilibrate for 4~ns. Data collection was performed on a subsequent 1~ns trajectory to calculate the local density, tetrahedrality, electric field, O-H bond frequency,  SFG spectra,  and hydrogen bonding statistics presented below. 

In all cases, five (5) replicas of each simulation were generated using random resampling of molecular placements to sample the surface attachment of the trace gasses and random re-seeding of atomic velocities prior to the first equilibration stage. In the sections that follow, all figures and data presented in tables represent the mean of the five replicas, and 95\% confidence intervals around these means were calculated using $\varepsilon  = 1.96 \sigma / \sqrt{N-1}$, where $N$ is the number of replicas, and $\sigma$ is the standard deviation of the sample mean. 

\section{Results}
\subsection{QLL and liquid film widths}
\begin{figure}
    \centering
    \includegraphics[width=\linewidth]{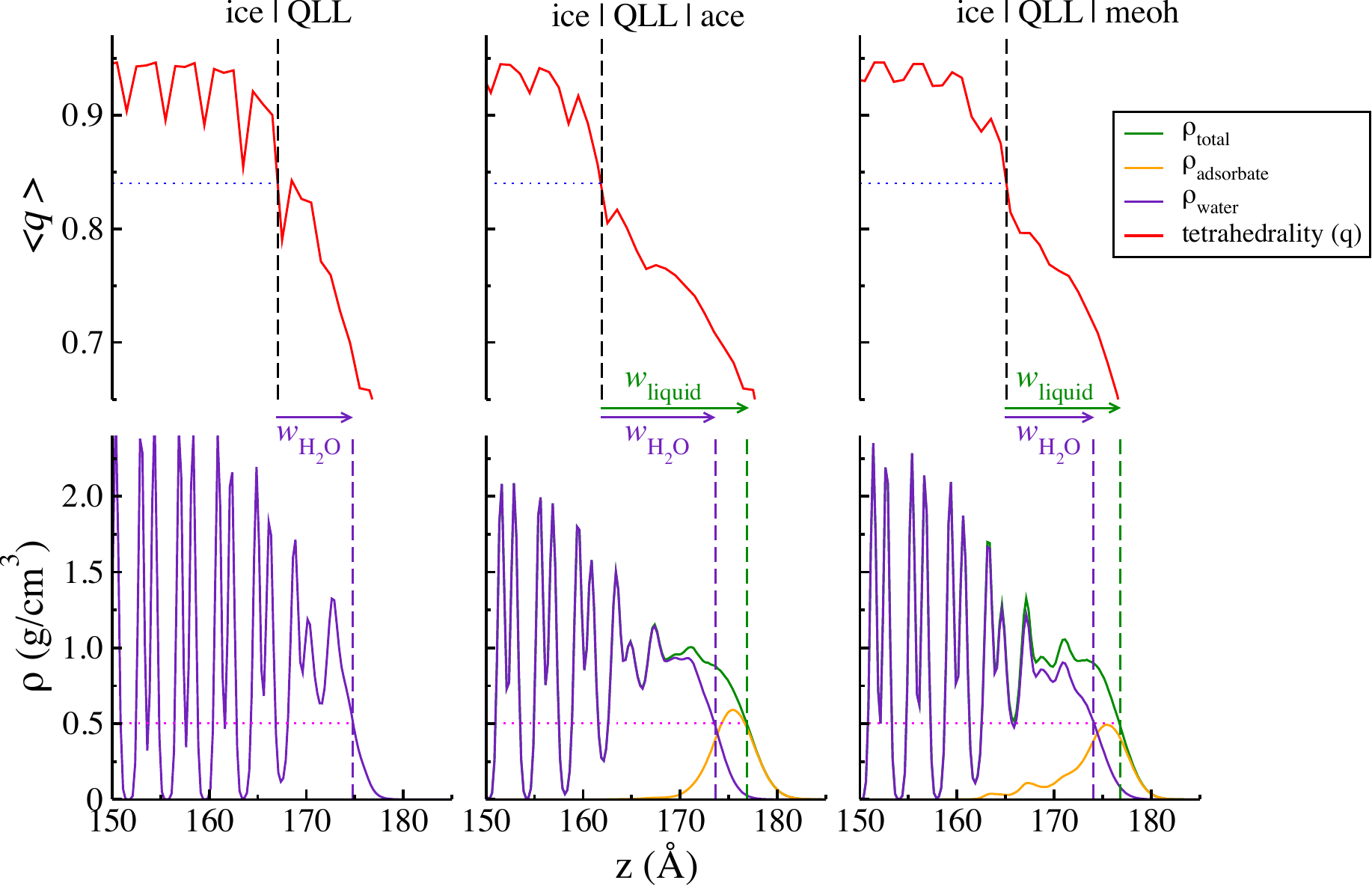}
    \caption{Local density (lower panels) and tetrahedral order parameter (upper panels) of the exposed ice-Ih basal facet (left), with adsorbed acetone (center), and with adsorbed methanol (right). These interfaces are simulated at 265~K. The position of the QLL/ice interface is referenced to the point where the tetrahedrality $q$ falls below 0.84. The dividing surfaces for the QLL and liquid layers are set where $\rho = 0.5 \mathrm{~g~cm}^{-3}$. We note methanol uptake in the outermost ice bilayer, indicated by the difference between the total density (green) and water density (purple) lines.}
    \label{fig:TetRhoAceMet265}
\end{figure}

\begin{figure}
    \centering
    \includegraphics[width=\linewidth]{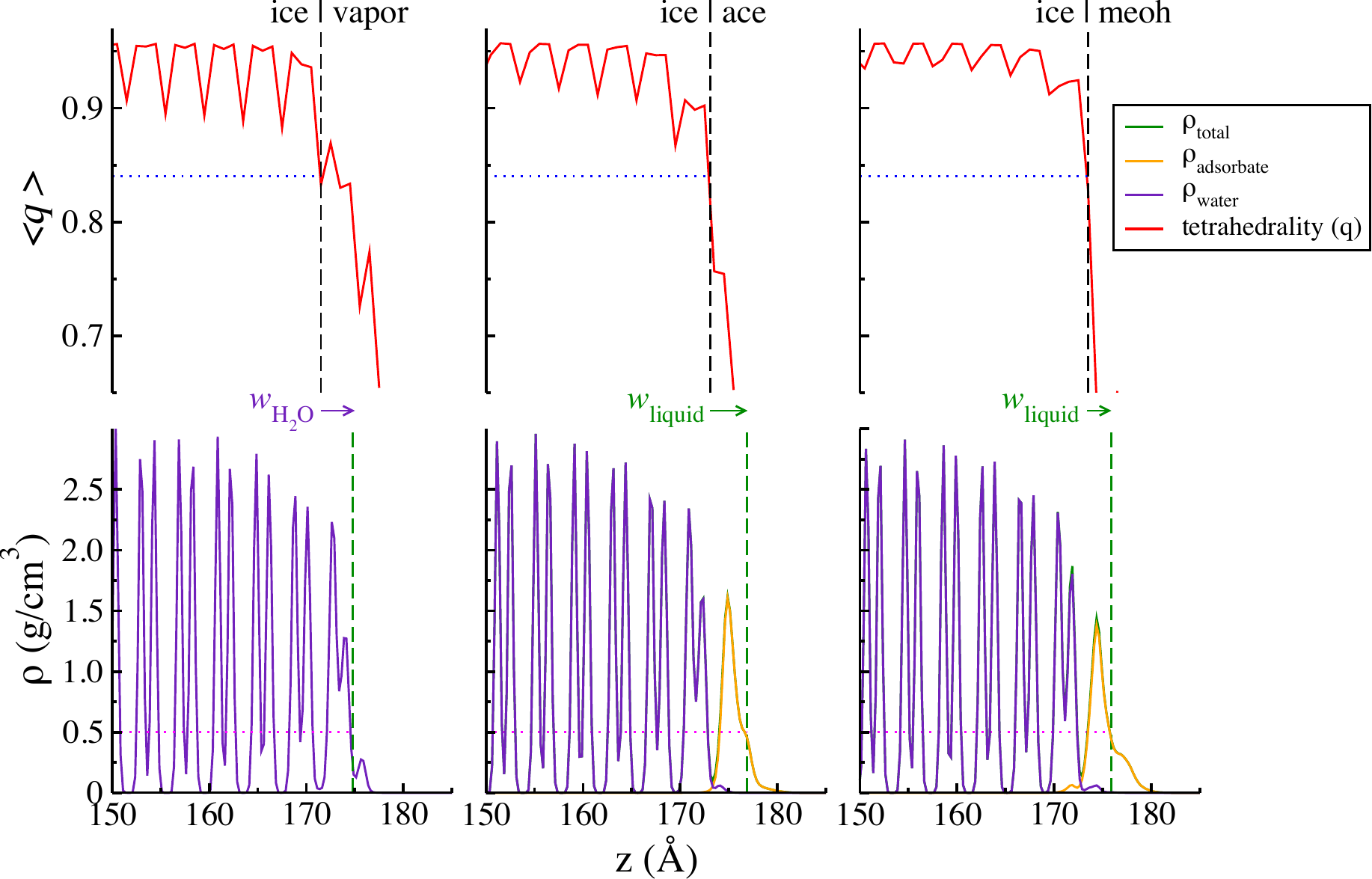}
    \caption{At 223~K, the local density vanishes between all TIP4P-Ice bilayers, indicating a persistent crystalline structure. This is evident even on the ice $|$ QLL $|$ vapor surface (left). For the surfaces with adsorbate molecules (center, right), the tetrahedrality $q$ only drops below 0.84 beyond the outer TIP4P-Ice bilayer, so no QLL is formed and the liquid layer consists mostly of the adsorbed gas.}
    \label{fig:TetRhoAceMet223}
\end{figure}

To assess the extent of adsorbate-enhanced melting, we estimated the widths of the quasi-liquid layer (QLL) via a process developed in previous work on TIP4P-Ice surfaces by Louden and Gezelter.\cite{Louden:2017aa} QLL widths are estimated by calculating the distance between two structural features, \textit{i.e.}, where the tetrahedral order parameter falls below a threshold $(q \le 0.84)$, defining the edge of the ice /  liquid interface, and where the local density falls below another threshold $(\rho \le 0.5 \mathrm{~g~cm}^{-3})$, which defines the edge of the liquid / vapor interface. The tetrahedral order parameter $(q)$ was first described by Errington and Debenedetti\cite{Errington2001}, and has been renormalized for under- and over-coordinated local structures.\cite{Louden:2017aa} It has been used extensively to describe ordering in ice / water interfaces and is largely a measure of local ordering in the water molecules surrounding a central water molecule. Tetrahedrality takes on large values ($q > 0.9$) in the ice crystal and falls ($q \sim 0.75$) in liquid water, although the exact value depends on both water model and temperature. Visual examples of how we have used tetrahedrality and density cutoffs to estimate liquid phase widths are shown in Figs. \ref{fig:TetRhoAceMet265} and \ref{fig:TetRhoAceMet223}.  Due to slight positional shifts of the ice crystal ($\sim 0.25$ \angstrom) in the 265~K systems, average density bins were aligned using the density well separating the central bilayer in the ice crystal. This was strictly for graphical representation in Fig. \ref{fig:TetRhoAceMet265} and did not alter the calculation of individual QLL widths. Average QLL and liquid film widths are provided in Table \ref{tab:LiqWidth}. 

We repeated this analysis for the lower temperature (223~K) simulations shown in \ref{fig:TetRhoAceMet223}, which closely match the temperature of the Yettapu \textit{et al.} experiments. At this temperature, the adsorbate-covered ice layers  preserve the crystalline tetrahedrality throughout the entire ice crystal, resulting in no observed QLL formation. Therefore, melting does not occur (${w_{\ce{H2O}}} \sim 0$) in the adsorbate systems and ${w_\mathrm{liquid}}$ at 223~K in Table \ref{tab:LiqWidth} reflects solely the thickness of the adsorbate monolayer. The bare ice / vapor interface exhibits more melting than the surfaces with adsorbates, and the threshold in the tetrahedral order parameter suggests that only the outer bilayer has enough disorder to form a QLL at this temperature. The liquid film widths are again provided in Table \ref{tab:LiqWidth}.

Both adsorbates studied here exhibited similar melting trends, as both appear to promote growth of the QLL at 265~K and suppress formation of the QLL at 223~K when compared to the bare ice / vapor surface. The estimated QLL widths of the 265~K basal ice facet fall within error estimates of a previous study on TIP4P-Ice at 265~K which predicted a QLL width of 7.2 angstroms.\cite{Louden:2017aa}

\begin{table}
    \centering
    \caption{Calculated Widths of the QLL and Liquid Layers (in \angstrom) for the basal ice facet with and without adsorbates. Values for structural widths $(w)$ are computed using the method shown in Fig. \ref{fig:TetRhoAceMet265}.  Note that at the lower temperature (223~K), the ice does not form a QLL, and the only liquid film is due to the adsorbate molecules. Uncertainties in the last digit are indicated with parentheses}
\label{tab:LiqWidth}
\begin{tabular}{ l | c c}
                \toprule
Adsorbate & $w_{\ce{H_2O}}$ & $w_{_{\mathrm{liquid}}}$ \\
\midrule
\multicolumn{3}{l}{T = 223~K} \\ \midrule
None & 1.8(5) & -- \\
Acetone & 0 & 3.65(9) \\
Methanol & 0 & 3.0(1) \\ \midrule
\multicolumn{3}{l}{T = 265~K} \\ \midrule
None & 7.9(8) & -- \\
Acetone & 12.4(7) & 15.6(7) \\
Methanol & 11(1)  & 14(1) \\
\bottomrule
\end{tabular}
\end{table}

We also note some differences in the distribution of adsorbates directly adjacent to the ice crystal. Methanol more readily permeates the QLL and integrates into the ice bilayers illustrated in Figure \ref{fig:systems}. In Figs.  \ref{fig:TetRhoAceMet265} and \ref{fig:TetRhoAceMet223}, differences between the water density (purple) and overall density (green) indicate some diffusion into the crystalline region. Fewer acetone molecules were observed crossing the QLL.   In table \ref{tab:AdosbratePermeability}, we provide numerical estimates of the number of molecules (per nm$^2$) of each adsorbate that penetrate into the QLL. We find nearly 4 times the concentration of methanol in this layer when compared with acetone. Adsorbate concentration in the ice is calculated by integrating the adsorbate number density up to the tetrahedrality cutoff ($q(z) \ge 0.84$), and in the QLL, by integrating from that point out to the Gibbs dividing surface between the QLL and the adsorbate layer ($\rho_{\ce{H2O}}(z) = 0.5$).  

\begin{table}
\begin{center}
\caption{ Average number of adsorbate molecules per nm$^2$ in the ice crystal and QLL in the 265~K systems. We find methanol (rather than acetone) permeates much more readily, with non-zero uptake in the ice crystal. }
\label{tab:AdosbratePermeability} 
\begin{tabular}{ l | c c }
                \hline
Adsorbate & Embedded in Ice (mol nm$^{-2}$) & Embedded in QLL (mol nm$^{-2}$)\\
\hline 
Acetone & 0 & 0.65(4) \\
Methanol & 0.01(1) & 2.5(1)\\
\hline
\end{tabular}
\end{center}
\end{table}

We find that methanol permeates the liquid layer of the high temperature systems significantly more than acetone does and is capable of integrating into the ice crystal itself.

\subsection{Frequency maps of surface O-H vibrations}

Because we do not observe melting in our ice/adsorbate models, we have looked for other explanations for the adsorbate-induced differences in the SFG spectra observed experimentally. To do so, we use electrostatic maps estimate O-H vibrational frequencies and hydrogen bonding statistics of the outermost water molecules. The aim is to correlate differences in local hydrogen bonding environments with the distribution of vibrational frequencies of the surface O-H bonds, which are indirectly measured by the SFG experiments.

\begin{figure}
    \centering
    \includegraphics[width=0.7\linewidth]{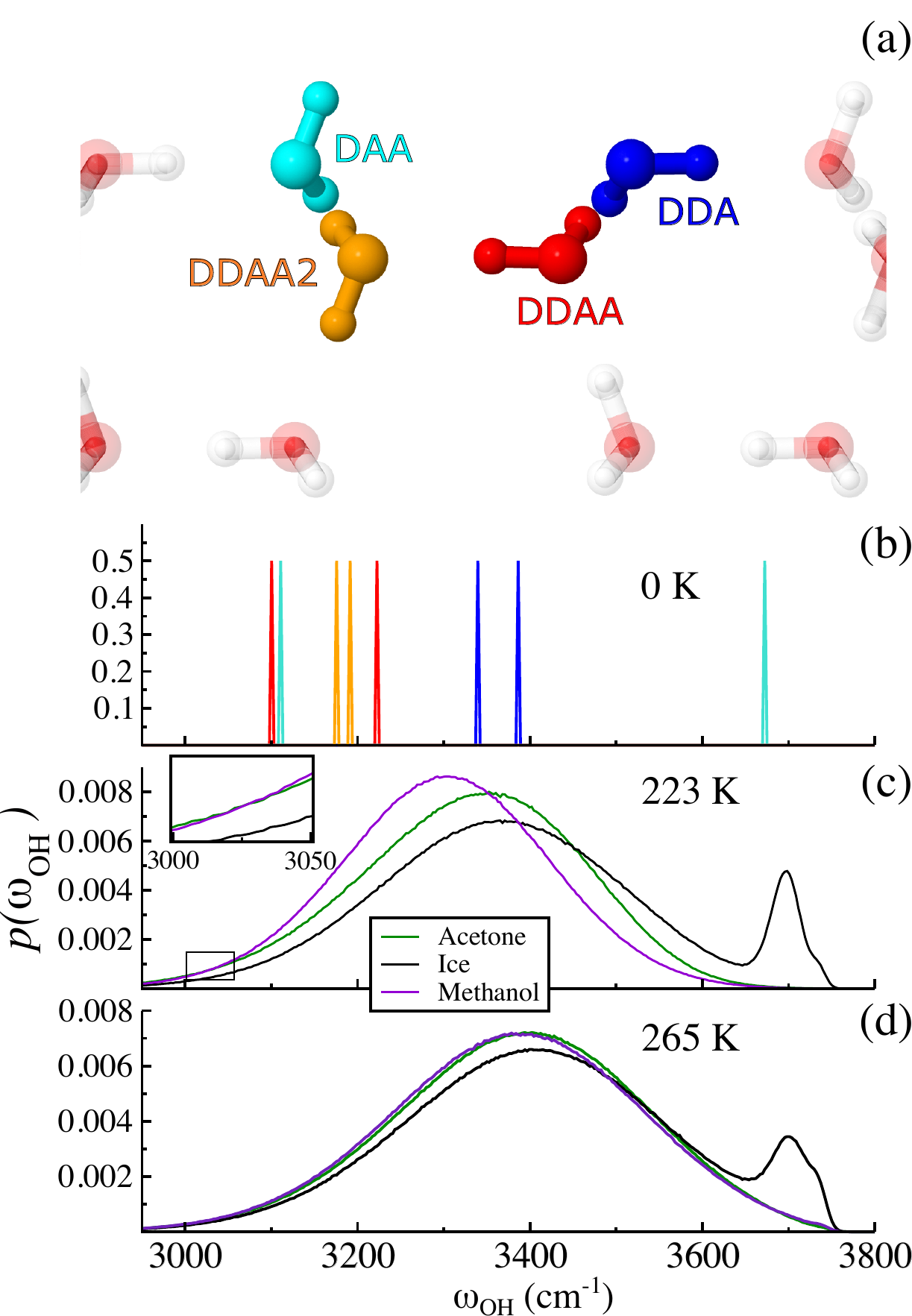}
    \caption{(a)~The four distinct molecules on the outer bilayer of the proton-striped basal $\{0001\}$ facet of ice that were used to generate the frequency data at 0~K. (b)~Eight O-H vibrational frequencies ($\omega_{~\ce{OH}}$) for the molecules identified in (a), computed using the electrostatic maps. (c)~Probability distributions of O-H bond frequencies in the outermost bilayer at 223~K, with the inset showing the crossover for the two adsorbates at low frequencies. (d)~Frequency contributions from the outer 50\% of the water film at 265~K.}
    \label{fig:4Panel}
\end{figure}

Corcelli, Lawrence, and Skinner pioneered the electrostatic mapping approach to estimating O-H vibrational lineshapes for dilute HOD in liquid water and \ce{D2O}.\cite{Corcelli:2004aa}  They observed a correlation between the local electric field projected along the O-H bond and the frequency of that bond. This approach was later expanded by Auer and Skinner to include non-linear contributions from the field, and the maps were further extended to produce transition dipoles and vibration-vibration couplings, specifically in the SPC/E water model.\cite{Auer:2008aa}  Grunenbaum \textit{et al.} later extended this work for the TIP4P water model.\cite{Gruenbaum:2013aa} Takayama \textit{et al.} have tested the transferability of the TIP4P maps to other TIP4P-derived models, including the TIP4P-Ice model utilized here.\cite{Takayama:2023aa}  Table S6 in the SI provides all of the electrostatic mapping parameters utilized in this work, both for the $p(\omega_\text{OH})$ plots in Fig. \ref{fig:4Panel} and for the calculations of the SFG spectra discussed below.

The theory for computing SFG spectra from these frequency maps was also developed largely by the Skinner group.\cite{Auer:2007aa,Auer:2008aa,Li:2010aa,Pieniazek:2011aa,Shi:2012aa,Ni:2015aa,Kananenka:2018aa} In this approach, classical molecular dynamics (MD) simulations are used along with the electrostatic maps to convert local electric fields into vibrational frequencies, transition dipoles, and vibrational couplings. These are used to construct a vibrational Hamiltonian in the O-H stretch and H-O-H bend overtone space.\cite{Gruenbaum:2013aa,Ni:2015aa,Kananenka:2018aa,Baiz:2020aa}  This Hamiltonian is then diagonalized, providing a time series of vibrational eigenstates. Vibrational spectra (IR, Raman, SFG) can then be computed from these eigenvalues and eigenvectors under the time-averaging approximation. This process is now well-defined, and the interpretations of calculated (and measured) SFG spectra of high frequency regions for water $(> 3400 \mbox{cm}^{-1})$ are relatively converged, while interpretation of the lower frequency range $(3000 - 3400 \mbox{~cm}^{-1})$ is still a matter of significant debate.\cite{Tang:2020aa}  

It was instructive to identify the particular water molecules  and O-H bonds at the surface of the ice interface that are most strongly impacted by the presence of the methanol or acetone adsorbates. We utilized the computed electric fields projected along the O-H bonds and then computed the distribution of O-H vibrational frequencies from the outermost water molecules present at the interface shown in blue in fig \ref{fig:systems}. We include in our sampling only the outermost bilayer of water molecules at 223~K and only the outer 50\% of the QLL at 265~K as water near or at the surface is most representative of the SFG signal. This is indicated via the set of the blue water molecules in Fig \ref{fig:systems}.

We start with a 0~K basal ice surface shown in the top panel of Fig. \ref{fig:4Panel}. This surface was constructed to expose the proton stripes first observed by Buch, \textit{et al.}\cite{Buch:2008aa} The SFG spectrum of the proton-striped surface has been characterized by Nojima, \textit{et al.}\cite{Nojima:2017aa}  In the outermost bilayer of water molecules, we observe 4 distinct hydrogen bonding environments. Using the notation of donors and acceptors,\cite{Smit:2017aa} we characterize these molecules as DAA, DDA, DDAA, and DDAA2. The primary difference between DDAA and DDAA2 molecules is whether they are hydrogen bonded to a DAA or DDA molecule residing on the surface. These four molecular environments yield eight distinct O-H vibrational frequencies shown in the subsequent panel. Note that the DAA molecule, or the water with a `free' O-H group, displays both high frequency $(3670 \mathrm{~cm}^{-1})$ and low frequency $(3110 \mathrm{~cm}^{-1})$ peaks, a phenomenon that is governed by the coupling between the two O-H vibrations and that was well characterized at water surfaces by Stiopkin \textit{et al.}\cite{Stiopkin:2011aa} 

In contrast, the oxygen-presenting surface molecule (DDA) produces two peaks (3340 and 3390 $\mathrm{~cm}^{-1})$ that are both within the broad $\sim 3400 \mathrm{~cm}^{-1}$ band for water. The DDAA molecule, which is directly hydrogen bonded to DDA, exhibits the lowest frequency near $3100 \mathrm{~cm}^{-1}$ and another peak at $3220 \mathrm{~cm}^{-1}$. The low frequency peak can be understood most directly in terms of the vibrational frequency maps, as the DDA provides a large negative charge directly along one of the DDAA O-H bonds without a surrounding solvent environment. The remaining fully-satisfied water molecule (DDAA2) is directly hydrogen bonded to DAA, exhibits two closely spaced peaks at 3175 and 3195 $\mathrm{~cm}^{-1}$ in this idealized geometry. These results generally agree with a similar study on a proton disordered ice crystal.\cite{Buch2007aa}

At higher temperatures, local disorder in the QLL obscures most of these details, except for the free O-H peak at $\sim 3700 \mathrm{~cm}^{-1}$ which is prominent in the bare ice simulations, but absent in the presence of either of the two adsorbates. At 223~K, there are significant differences in the broad peak position and shape that depend on the identity of the adsorbates. Ice with adsorbed acetone exhibits a broad distribution, closely aligned with the DDA frequencies, while the methanol adsorbate produces a narrower, red-shifted peak, centered near $\sim 3300 \mathrm{~cm}^{-1}$. Due to the narrower distribution of the water O-H frequencies with adsorbed methanol, we also observe a crossover point near $3025 \mathrm{~cm}^{-1}$ where ice with the adsorbed acetone exhibits larger populations of the low frequency O-H bonds relative to the methanol adsorbate. This is shown in the inset graph of Figure \ref{fig:4Panel}.

Our hypothesis is that this is largely due to methanol's ability to satisfy hydrogen bonds of both under-coordinated sites (DAA \& DDA), while acetone can only satisfy the dangling proton on DAA. Indeed, the largest difference between the methanol and acetone adsorbate frequency distribution is directly in line with the DDA peaks at 3340-3390 $\mathrm{~cm}^{-1}$ 

Differences in the distributions at lower frequencies do exist between the two adsorbates, and this can be understood in the changing electrostatic environment of the DDAA molecules when DDA is participating in a hydrogen bond with an adsorbate molecule. The electrostatic contribution of the H-bonded methanol on DDA can shift the low frequency contribution of DDAA $(3100 \mathrm{~cm}^{-1})$ higher in frequency space. Additionally, the narrower distribution of methanol is indicative of a more uniform distribution of electric fields surrounding the O-H bonds of the interfacial water molecules.

In the SI, we have included additional bilayers in calculating these frequency distributions. We find that the inclusion of additional bilayers red-shifts the peak center for the acetone and ice samples, bringing them into alignment with the central frequency of the methanol sample.
This suggests that the outermost water bilayer with a methanol adsorbate has more `bulk-like' electrostatic environments that drive O-H vibrations to lower frequency regions.

\subsection{Calculated SFG spectra}
\begin{figure}
    \centering
    \includegraphics[width=\linewidth]{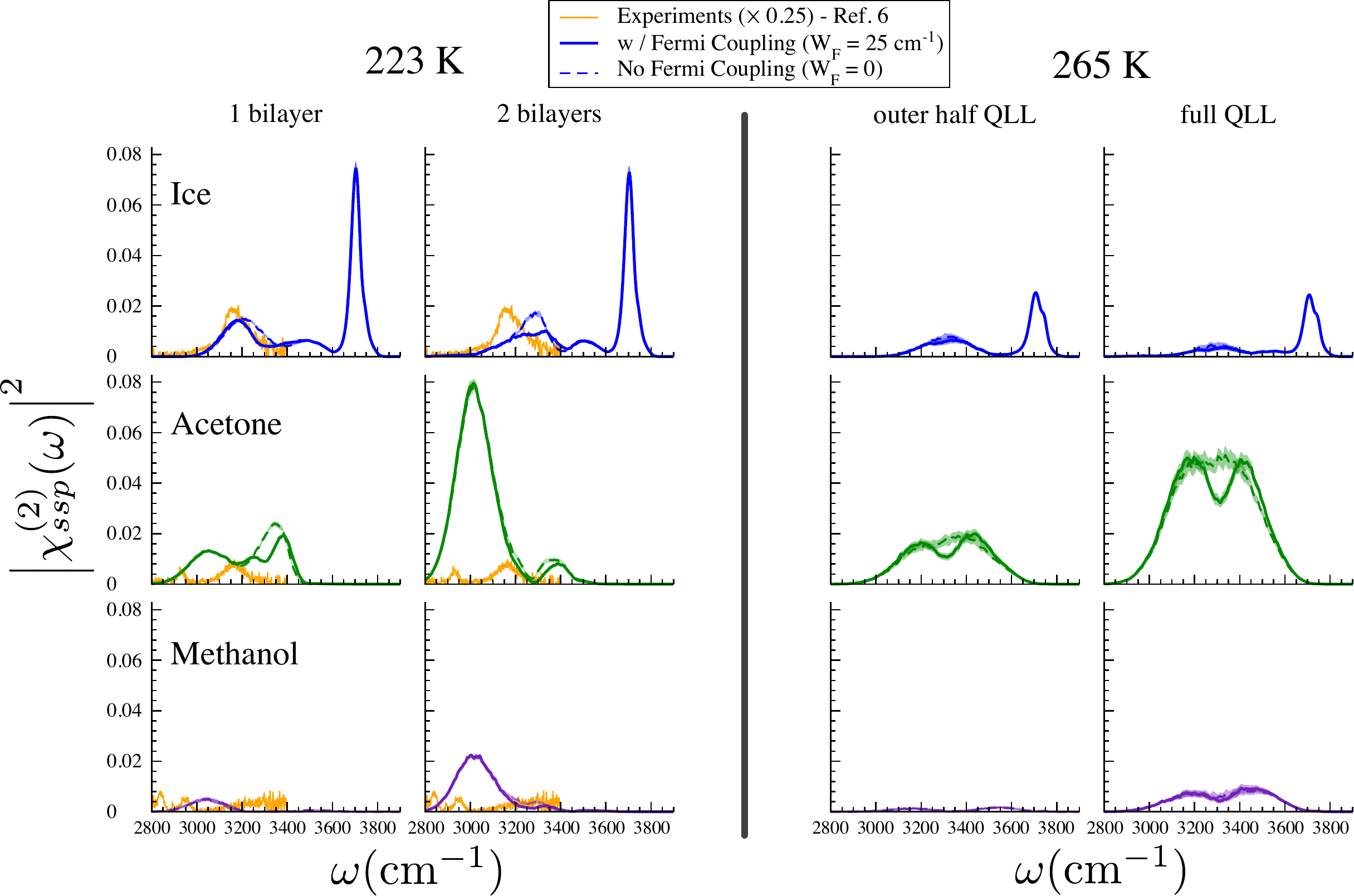}
    \caption{ O-H vibrational SFG signals $|\chi_{ssp}^2(\omega)|^2$ calculated using electrostatic maps and the time averaging approximation.  Top: for bare ice interfaces. Middle: for surfaces with acetone adsorbates. Bottom: for surfaces with methanol adsorbates. Spectra were calculated using 1 and 2 bilayers of the ice (at 223~K) or using half and full widths of the quasi-liquid layer (at 265~K). Signals were computed using Fermi coupling between the H-O-H bend overtones and O-H stretches (solid lines, $W_F = 25 \text{~cm}^{-1}$) and also without the Fermi coupling (dashed lines, $W_F = 0$). Note that the upper frequency range of the ice surface experiments in Ref. \citenum{Yettapu:2025aa} was 3400 $\text{cm}^{-1}$  }
    \label{fig:TAA}
\end{figure}

To make explicit connections with the experiments, we have used electrostatic maps to construct the exciton Hamiltonian in O-H stretch fundamentals
together with the H-O-H bend overtones (details are provided in the SI). When diagonalized, this Hamiltonian provides a set of eigenvalues $\{\omega_a\}$ and eigenstates $\{\mathbf{V}_{a}\}$, each with a transition dipole and transition polarizability constructed from the local modes $\{i\}$ in the exciton basis,
\begin{equation}
  \mu^a_r = \sum_i V_{ia}\,\mu_{r,i}, \qquad
  \alpha^a_{pq} = \sum_i V_{ia}\,\alpha_{pq,i},
\end{equation}
$\mu_{r,i}$ is the $r$-component of the transition dipole of local mode $i$ and $\alpha_{pq,i}$ is the $pq$-component of its bond polarizability tensor, with the Cartesian indices $p,q,r \in \{x,y,z\}$ labeling the laboratory-frame components. Each choice
of $(p,q,r)$ corresponds to a measurable polarization combination.
Under the time averaging approximation, the second-order susceptibility can be constructed as a sum of complex Lorentzians,\cite{Auer:2007aa,Auer:2008aa}
\begin{equation}
  \chi^{(2)}_{pqr}(\omega) \propto
    \left\langle \sum_a
      \frac{\alpha^a_{pq}\,\mu^a_r}{\omega - \omega_a - i\Gamma}
    \right\rangle,
    \label{eq:susceptibilityTAA}
\end{equation}
where the angle brackets denote an average over MD configurations.  The  half-width, $\Gamma$, in the Lorentzian corresponds to a dephasing time $T_2$ which has contributions from the vibrational lifetime and from pure dephasing. Typical vibrational lifetimes in water are estimated to be $200-700$ fs, so $\Gamma = 5 \mathrm{~cm}^{-1} ~~(T_2 \sim 1~\mathrm{ps})$ is a reasonable choice in ice, and this value was utilized in calculating the spectra in Fig. \ref{fig:TAA}. 
For the $ssp$ polarization combination (with the interface normal along $z$), the
relevant response is
\begin{equation}
  \chi^{(2)}_{ssp}(\omega) \propto \text{Im}\left[\frac{1}{2}\bigl(\chi^{(2)}_{xxz}(\omega) + \chi^{(2)}_{yyz}(\omega)\bigr)\right],
\end{equation}
where the two equivalent in-plane components are averaged to improve
statistics.\cite{Kananenka:2018aa} Because the SFG response depends on the direction of the surface normal, some care with sign conventions is needed to average contributions from the top and bottom surfaces.

In Fig. \ref{fig:TAA}, we show the calculated O-H vibrational SFG signal, $|\chi_{ssp}^2(\omega)|^2$, for the same molecules and surfaces used in Fig. \ref{fig:4Panel}.  The 223 K spectrum represents the averages of both faces of the ice crystal from five independent simulations.  At 265~K, the spectra come from sampling half of the QLL depth from 5 independent simulations (each with two exposed interfaces), which provides data only from the water molecules close to the liquid/vapor interface. We also include calculations of an additional ice bilayer in the 223~K systems and the full QLL in the 265~K systems. These larger sampling regions increase the signal from bulk-like regions and do not align as closely with the experimental SFG data from Ref. \citenum{Yettapu:2025aa}.

For the bare ice at 223~K, the simulated $|\chi_{ssp}^2(\omega)|^2$ includes three distinct spectral features.  The dominant peak at $3700 \mathrm{~cm}^{-1}$ is the signature of the free O-H groups, which includes both surface TIP4P-Ice and a higher frequency shoulder due to TIP4P-Ice molecules that have drifted into the vapor layer. We attribute the feature at 3500 $\mathrm{cm}^{-1}$ to molecules that are fully-participating in the Hydrogen bonding network. The lowest frequency feature, centered near 3180 $\mathrm{~cm}^{-1}$ is influenced by the Fermi coupling between O-H stretches and the H-O-H bend overtone, and likely corresponds to collective motion in the bilayer planes. The surface with adsorbed acetone  features three distinct peaks, with one near $3400 \mathrm{~cm}^{-1}$ and overlapping features near 3040 and 3240 $\mathrm{~cm}^{-1}$. Here, the Fermi coupling strongly influences the two highest frequency features. The surface with methanol exhibits only 2 broad peaks, one near 3040 $\mathrm{~cm}^{-1}$ and a second, weaker feature at 3500 $\mathrm{cm}^{-1}$.

These computed spectra provide one hypothesis for the disappearing experimental feature near 3200 $\mbox{cm}^{-1}$. As suggested by the distribution of O-H frequencies, this could be a contribution from the DDAA molecules as DDA becomes hydrogen bound, either through adsorption of methanol or via surface melting. If the under-accepting water participates in a new hydrogen bond, the local field at DDAA is significantly altered. The acetone adsorbate largely preserves response in this frequency range, lending some support to this hypothesis. 

For the bare ice at 265~K, the simulated $|\chi_{ssp}^2(\omega)|^2$ again exhibits a dominant free O-H feature at $3708 \mathrm{~cm}^{-1}$. This signal is not present in either of the adsorbate systems. There is an additional broad peak centered on $3322 \mathrm{~cm}^{-1}$. Both methanol and acetone exhibit much broader distributions of signal relative to the calculated 223~K signal. We note that the features from the methanol / QLL / ice surface are blue shifted relative to their appearance in the lower-temperature methanol / ice surface. Assigning specific contributions of the interfacial water molecules to individual peaks in the low frequency composite SFG signal remains a challenge.

\subsection{Hydrogen bonding statistics}
To characterize the populations of under- and fully-coordinated water molecules in the outer layer, we calculated the number of hydrogen bonds in which these molecules participated and classified the molecules by whether they were hydrogen bond donors or acceptors. We again restrict data collection to the same criterion as the frequency maps, as those water molecules would be most strongly represented in the SFG spectra. The sampled molecules are shown in blue in Fig. \ref{fig:systems}. 

\begin{table}
    \centering
    \caption{Hydrogen bonding statistics for outermost water molecules. For the 223~K systems, this includes all water molecules in the outermost bilayer. For the 265~K systems, this includes water in the outer half of $w_{\ce{H2O}}$. Molecules that under-donate ($n_D = 1$) are identified as DAA water molecules, while those that under-accept ($n_A=1$) are identified as DDA.}
\label{tab:FreeHBond}
\begin{tabular}{ l | l l | l l | l l }
                \toprule
 \multirow{2}{*}{Interface} & \multicolumn{2}{c|}{Hydrogen Bonding} & \multicolumn{2}{c|}{Donating} & \multicolumn{2}{c}{Accepting} \\ \cline{2-3} \cline{4-5} \cline{6-7} 
  & nHB $<$ 4 & nHB~=~4 & nD~=~1  & nD~=~2  & nA~=~1 & nA~=~2 \\
\midrule
\multicolumn{7}{l}{T = 0K} \\ \midrule
 $\{0 0 0 1 \}$ & 0.5 & 0.5  & 0.25 & 0.75 & 0.25 & 0.75\\ \midrule
\multicolumn{7}{l}{T = 223~K} \\ \midrule
ice\textbar vapor & 0.351(3)  & 0.643(2)  & 0.185(2) & 0.813(2)  & 0.196(2) & 0.798(2) \\
ice\textbar ace & 0.253(2) & 0.745(2) & 0.0059(5) & 0.9938(5)  & 0.248(2) & 0.750(2) \\
ice\textbar meoh & 0.093(2) & 0.905(2) & 0.0052(2) & 0.9945(2)  & 0.089(2) & 0.909(2) \\ \midrule
\multicolumn{7}{l}{T = 265~K} \\ \midrule
QLL & 0.24(2) & 0.73(2) & 0.125(8) & 0.869(8) & 0.15(1) & 0.82(1) \\
QLL\textbar ace & 0.216(5) & 0.752(5) & 0.059(1)  & 0.937(1)  & 0.174(4)  & 0.794(4)  \\
QLL\textbar meoh & 0.19(1) & 0.78(1) & 0.056(3) & 0.940(3)  & 0.15(1) & 0.82(1) \\

\bottomrule
\end{tabular}
\end{table}

The primary contrast between the adsorbate systems at 223~K is evident in the hydrogen bond acceptor populations. Adsorption of acetone preserves populations of hydrogen bond under-acceptors (DDA) described previously in Fig \ref{fig:4Panel}, in nearly identical number to the ideal 0~K basal crystal. This occurs because acetone accepts hydrogen bonds from the DAA at the outermost ice bilayer, but cannot act as a hydrogen bond donor to the DDA molecules. In contrast, methanol interacts equitably, fully satisfying hydrogen bonds in the outermost ice bilayer. It is also capable of integrating into the ice crystal. Both systems severely deplete the free O-H populations represented by the under-donating statistics in Table \ref{tab:FreeHBond} at 223~K. As we move to higher temperatures, all 3 systems behave similarly in terms of hydrogen bonding.   We note that a similar table providing counts of molecules in each of these configurations is supplied in the SI.

\section{Discussion \& Conclusions}
We constructed our simulations to model the experimental setup of Yettapu \textit{et al.} and to interrogate the surface melting and hydrogen bonding characteristics that could contribute to the observed SFG signal shift between acetone and methanol at the peak originating at $\sim 3170 \mathrm{~cm}^{-1}$. Identical proton ordered striped basal ice crystals were simulated with a vacuum layer and with monolayers of methanol and acetone at 223~K and 265~K, the former matching the temperature of the sum frequency generation (SFG) experimentation and the latter producing a liquid-like environment for the adsorbates on the QLL. 

Using structural order parameters, \textit{i.e.,} tetrahedral order and local density, to define the boundaries between the ice and the QLL and also between the QLL and the vapor, we did not observe any differences in the degree of melting between interfaces with adsorbed acetone or methanol at either temperature. Here, the degree of melting is taken from the thickness of the QLL itself. Both systems exhibited enhanced melting at 265~K and suppressed melting at 223~K relative to the bare ice / vapor interfaces at the same temperatures. Any observed melting in our simulations is therefore not sufficient to explain the difference in the observed SFG signal. 

Using electrostatic mapping, we characterized the frequencies of the O-H bonds in the outermost bilayer of the ice crystal that would be most directly impacted by the adsorbates, focusing on the O-H bonds that were near $3170 \mathrm{~cm}^{-1}$. Of the O-H bonds, we find that the lowest O-H bond frequencies are present on the fully coordinated water molecule (DDAA) bound to the under-donating (DDA). In addition, the low frequency partners of the free O-H (DAA) exhibit frequencies near $\sim 3100 \mathrm{~cm}^{-1}$. At 223~K, we observe differences in the broad peak from the surface water molecules that depend strongly on the adsorbate. The broad feature with adsorbed acetone closely aligns with the DDA peaks at $3340 - 3390 \mathrm{~cm}^{-1}$, while adsorbed methanol produces a narrower feature with a peak that has red shifted relative to the ice / vapor interface. However, the lowest frequency bonds ($< 3025 \mathrm{~cm}^{-1}$) are more prevalent in the acetone system. The prominent differences in the hydrogen bond statistics reinforce this observation, notably for the molecules that are accepting only one hydrogen bond (DDA). Acetone preserves ice-like populations of DDA surface molecules, while methanol is able to incorporate into the ice crystal and can satisfy the hydrogen bonding disparity, converting DDA molecules to DDAA. As both adsorbates act as hydrogen bond acceptors, the populations of free O-H (DAA) bonds are depleted and are much less prevalent than in the bare ice.

Our results suggest another possible source of the observed changes in SFG signal at $\sim 3170 \mathrm{~cm}^{-1}$ rather than surface melting. In the acetone simulations, the DDAA low frequency O-H bond remains largely unperturbed, which could be responsible for the red shift seen in the experiment. In contrast, the methanol adsorbates produce full coordination and depletion of the under-coordinated waters in the outer bilayer, which may account for the blue-shift of the low frequency DAA and DDAA O-H bonds in the SFG signal. In the language of the electrostatic frequency maps, this can be attributed to the electric field around this pair of water molecules, which is impacted by a direct hydrogen bond between methanol and the under-coordinated water molecules.

So far, the simulations and the experimental SFG signals do not seem to have fully converged on an explanation. 
 We note two possible sources for this disagreement. First, the partial charges utilized in our adsorbate models for methanol and acetone may not provide the electric fields necessary to reproduce the correct SFG spectra.  Additionally, the time-averaging approximation to the SFG spectra neglects motional
narrowing.  Because each configuration is treated as static, the
computed linewidths are governed by the inhomogeneous distribution of
eigenfrequencies and the phenomenological width, $\Gamma$, ignoring the averaging of frequency fluctuations over the vibrational
lifetime. This treatment also misses any frequencies that would appear in the O-H stretch due to coupling with C-H motion in the adsorbates. In future work,
explicit propagation of the exciton coherences and the use of a time-correlation function formalism~\cite{Kananenka:2018aa} will allow for a more complete picture of the SFG signals. This approach should provide a better match with experimental observables. 

\section{Data and Software Availability}
The data used in this study (force field parameters, initial configurations, extracted data, graphs, scripts, and meta-data) are available at DOI: \href{https://doi.org/10.5281/zenodo.20571172}{10.5281/zenodo.20571172}  All simulations utilized the OpenMD molecular dynamics engine (See \href{https://github.com/OpenMD/OpenMD}{github.com/OpenMD/OpenMD}),\cite{Drisko:2024aa} which is available under a BSD 3-clause license . All analysis tools utilized here have been built into OpenMD. 

\begin{acknowledgement}
Support for this project was provided by the National Science Foundation under grant CHE-1954648. Computational time was provided by the Center for Research Computing (CRC) at the University of Notre Dame.
\end{acknowledgement}

\begin{suppinfo}
Force field parameters, including atom type properties, bond, bend, torsion, and inversion function parameters. Electrostatic map parameters for calculation of O-H frequencies. Procedure for computing SFG spectra from electrostatic maps. Complete Hydrogen bonding statistics. O-H vibrational frequency distributions in additional ice bilayers.
\end{suppinfo}

\section{Author Contributions}
This work was made available through contributions from both authors. Both authors have approved the final version of the manuscript. B.M.H and J.D.G. conceived and designed the simulations; B.M.H. performed the simulations; B.M.H. and J.D.G. analyzed the data; B.M.H and J.D.G. wrote and edited the manuscript. J.D.G. secured financial support for the research.

\section{Notes}
The authors declare no competing financial interests.

\newpage

\bibliography{CiteKey}

\end{document}


\section{Forcefield parameters}
Force field parameters for methanol (\ce{CH3OH}) and acetone (\ce{(CH3)2CO}) were taken from the Generalized Amber Force Field (GAFF)\cite{Wang:2004aa} with charges on the methanol adapted from the work of Fennell, Wymer, and Mobley, which provides accurate descriptions of solvation free energies for alcohols.\cite{Fennell:2014aa}  Partial charges on the acetone atoms were obtained using the Austin Model 1 -- with bond charge corrections (AM1-BCC) model,\cite{Jakalian:2000aa,Jakalian:2002aa,ACPYPE}
Rigid TIP4P-Ice water molecules\cite{Abascal:2005aa} were used as models for the solid ice and the QLL layer.
The damped shifted force (DSF) kernel by Fennell and Gezelter was used for all long-range electrostatic interactions with a damping parameter ($\alpha$) of 0.18 \AA$^{-1}$.\cite{Fennell:2006aa}   Table \ref{tab:atypes} provides the mass, charge, and non-bonded (Lennard-Jones) parameters for the species used in this work. The Lorentz-Berthelot combining rules,
\begin{align*}
\sigma_{ij} &= \frac{\sigma_{i} + \sigma_{j}}{2} \\ 
\varepsilon_{ij} &= \sqrt{\varepsilon_i \varepsilon_j}
\end{align*}
were employed for Lennard-Jones interactions between different atom types. Atom pairs within the same rigid body, as well as atom pairs involved in bonding (1-2) or bending (1-3) potentials did not have dispersion or electrostatic interactions. However, atom pairs involved in torsion (1-4) potentials had the standard GAFF2 scaling of electrostatic ($5/6 = 0.8\bar{3}$) and van der Waals ($1/2 = 0.5$) interactions.

\begin{table}
\begin{center}
\caption{Mass, charge, and Lennard-Jones parameters used in these simulations} \label{tab:atypes} 
\bibpunct{}{}{,}{n}{,}{,}
\begin{tabular}{ r | r r r r r}
                \hline
AtomType & Mass (amu) & Charge ($e$) & $\varepsilon$ (kcal/mol) & $\sigma$ (A) & Source \\
\hline 
\multicolumn{5}{l}{Acetone} & Refs. \cite{Wang:2004aa}, \cite{Jakalian:2000aa} and \cite{Jakalian:2002aa} \\ \hline
C & 12.01 & 0.561102 & 0.078 & 3.634867 & \\
C3 & 12.01 & -0.204100 & 0.078 & 3.634867 & \\
HC & 1.008 & 0.063033 & 0.0208 & 2.6001 & \\ 
O & 16.0 & -0.531100 & 0.120 & 3.029056 & \\ \hline
\multicolumn{5}{l}{Methanol} & Refs. \cite{Wang:2004aa} and \cite{Fennell:2014aa}\\ \hline
C3 & 12.01 & 0.14110 & 0.109400 & 3.39967\\
HC & 1.008 & 0.03470 & 0.0157 & 2.47135\\
OH & 16.0 & -0.72398 & 0.202074 & 3.21990\\ 
HO & 1.008 & 0.47878 & 0.0000 & $1 \times 10^{-6}$ \\ \hline
\multicolumn{5}{l}{TIP4P-Ice} & Ref. \cite{Abascal:2005aa}\\ \hline
O & 16.0 & 0.0 & 0.210842 & 3.1668 & \\
H & 1.008 & 0.5897 & 0.0 & 0.0 & \\
EP & 0.0 & -1.1794 & 0.0 & 0.0 & \\
\hline
\end{tabular}
\end{center}
\end{table}

In Table \ref{tab:bond}, harmonic bonds are described by
\begin{equation}
    V_\text{bond}(r) = k_b (r - b_0)^2
\end{equation}
where $k_b$ is the force constant and $b_0$ is the equilibrium bond length. Water is simulated as a rigid body so no harmonic bond parameters are needed. The base atom types are shown here when applicable, so C refers to all carbon atoms that are not accounted for in the other listed harmonic bond parameters (e.g., C3) and O refers to all oxygen atoms that are not otherwise listed (e.g., OH). 

\begin{table}
\bibpunct{}{}{,}{n}{,}{,}
    \centering
    \caption{Harmonic Bond Parameters}
\label{tab:bond}
\begin{tabular}{ c c | c c c}
                \hline
Atom1 & Atom2 & $b_0$ (\AA) & $k_b$ (kcal/mol) & Source \\
\hline
\multicolumn{4}{l}{Acetone} & Ref. \cite{Wang:2004aa}\\ \hline
C & O & 1.218 & 652.57 & \\
C & C3 & 1.524 & 243.22 & \\
C3 & HC & 1.097 & 375.92 & \\
\hline
\multicolumn{4}{l}{Methanol} & Ref. \cite{Wang:2004aa} \\ \hline
OH & HO & 0.973 & 563.51 & \\
C3 & OH & 1.423 & 293.40 & \\
C3 & HC & 1.097 & 375.92 & \\
\hline
\end{tabular}
\end{table}

In Table \ref{tab:bend}, harmonic bends are in the form
\begin{equation}
V_{\text{bend}}(\theta) = k_\theta (\theta - \theta_0)^2
\end{equation}
where $\theta$ describes the angle between bonded atoms of type $i-j-k$, where $j$ is the central atom.

\begin{table}
\bibpunct{}{}{,}{n}{,}{,}
    \centering
    \caption{Harmonic Bend Parameters}
\label{tab:bend}
\begin{tabular}{ c c c | c c c }
                \hline
 
Atom1 & Atom2 & Atom3 & $\theta_0~(^\circ)$ & $k_\theta \mathrm{~(kcal~mol^{-1}~rad^{-2})} $ & Source\\
\hline
\multicolumn{5}{l}{Acetone}  & Ref. \cite{Wang:2004aa}\\ \hline
HC & C3 & HC & 107.580 & 38.960 &\\
HC & C3 & C & 110.360 & 47.475 &\\
C3 & C & O & 123.200 & 84.552/c &\\
\hline
\multicolumn{5}{l}{Methanol} & Ref. \cite{Wang:2004aa} \\ \hline
HC & C3 & HC & 107.580 & 38.960 &\\
HC & C3 & OH & 109.500 & 62.756 &\\
C3 & OH & HO & 107.260 & 49.027 &\\
\hline
\end{tabular}
\end{table}

In Table \ref{tab:torsions}, CHARMM-style torsions take the form 
\begin{equation}
V_{\text{torsion}}(\phi) = \sum_n K_n (1 + \cos(n\phi - \delta_n))
\end{equation}
where $K_n$ is the amplitude and $\delta_n$ is the phase angle corresponding to a cosine contribution with periodicity $n$. 

\begin{table}
\bibpunct{}{}{,}{n}{,}{,}
    \centering
    \caption{Torsion parameters}
\label{tab:torsions}
\begin{tabular}{ c c c c | c c c c}
                \hline
 
Atom1 & Atom2 & Atom3 & Atom4 & $K_n \mathrm{(kcal~mol^{-1})}$ & $n$ & $\delta (^\circ)$ & Source \\
\hline
\multicolumn{7}{l}{Acetone} & Ref. \cite{Wang:2004aa}\\ \hline
HC & C3 & C & O & 0.830 & 1 & 0 &\\
HC & C3 & C & O & 0.040 & 3 & 180 &\\
HC & C3 & C & C3 & 0.000 & 2 & 0 &\\
\hline
\multicolumn{7}{l}{Methanol} & Ref. \cite{Wang:2004aa}\\ \hline
HC & C3 & OH & HO & 0.500 & 3 & 0  &\\ \hline
\end{tabular}
\end{table}

For inversions centered on $\mathrm{sp}^2$-hybridized atoms with exactly three satellite atoms, the Amber\cite{AmberTools} improper potential is utilized, which has the form,
\begin{equation}
 V_{\textrm{inversion}} (\omega_{ijkl}) = \frac{v}{2} \left[1 - \cos \left( 2 (\omega_{ijkl} - \omega_{ijkl}^0)\right)\right]
    \label{eq:inversion}
\end{equation}
where $\omega_{ijkl}$ is an improper torsion angle with the central atom in position 3 of a standard torsion, and the satellite atoms are in positions 1, 2, and 4. In table \ref{tab:inversion}, the central atom of an inversion is listed first.

\begin{table}
\bibpunct{}{}{,}{n}{,}{,}
    \centering
    \caption{Inversion Parameters}
\label{tab:inversion}
\begin{tabular}{ r | l l l l | c c r}
                \hline
 
Molecule & Central Atom & Atom2 & Atom3 & Atom4 & $\omega_0 (^\circ)$ & $v$ (kcal/mol) & Source\\
\hline
Acetone & C & C3 & C3 & O & 0 & 10.5 & Ref. \cite{Wang:2004aa}\\ \hline

\end{tabular}
\end{table}

\subsection{Calculation of the SFG signals}
Corcelli, Lawrence, and Skinner pioneered the electrostatic mapping approach to estimating O-H vibrational lineshapes for dilute HOD in liquid water and \ce{D2O}.\cite{Corcelli:2004aa}  They observed a correlation between the local electric field projected along the O-H bond and the frequency of that bond. This approach was later expanded by Auer and Skinner to include non-linear contributions from the field, and the maps were further extended to produce transition dipoles and vibration-vibration couplings, specifically in the SPC/E water model.\cite{Auer:2008aa}  Grunenbaum \textit{et al.} later extended this work for the TIP4P water model.\cite{Gruenbaum:2013aa} Takayama, \textit{et al.} have tested the transferability of the TIP4P maps to other TIP4P-derived models, including the TIP4P-Ice model utilized here.\cite{Takayama:2023aa}  Table \ref{tab:maps} provides all of the electrostatic mapping parameters utilized in this work, both for the $p(\omega_\text{OH})$ plots in Fig. 4 in the paper and for the calculations of the SFG spectra presented in Fig. 5.

The theory for computing SFG spectra from electrostatic frequency maps was also developed largely by the Skinner group.\cite{Auer:2007aa,Auer:2008aa,Li:2010aa,Pieniazek:2011aa,Shi:2012aa,Ni:2015aa,Kananenka:2018aa} In this approach, classical molecular dynamics (MD) simulations are used along with the electrostatic maps to convert local electric fields into vibrational frequencies, transition dipoles, and vibrational couplings. These are used to construct a vibrational Hamiltonian in the O-H stretch and H-O-H bend overtone space.\cite{Kananenka:2018aa,Gruenbaum:2013aa,Baiz:2020aa}  This Hamiltonian is then diagonalized, providing a time series of vibrational eigenstates. Vibrational spectra (IR, Raman, SFG) can then be computed from these eigenvalues and eigenvectors under the time-averaging approximation. This process is now well-defined, and the interpretations of calculated (and measured) SFG spectra of high frequency regions for water $(> 3400 \mbox{cm}^{-1})$ are relatively converged, while interpretation of the lower frequency range $(3000 - 3400 \mbox{~cm}^{-1})$ is still a matter of significant debate.\cite{Tang:2020aa}  

\begin{table}[htbp]
  \centering
  \caption{Electrostatic spectroscopic map parameters used for the
    TIP4P water models. The O-H stretch maps are those of
    Gruenbaum \textit{et al.}\cite{Gruenbaum:2013aa} and the H-O-H bend
    overtone maps are those of Ni and Skinner.\cite{Ni:2015aa} The same
    parameters are applied to TIP4P-Ice, following the transferability
    demonstrated by Takayama \textit{et al.}\cite{Takayama:2023aa} The
    projected electric field $E$ and all matrix elements are in atomic
    units; frequencies are in \mbox{cm}$^{-1}$.}
  \label{tab:maps}
  \begin{tabular}{l l l}
    \toprule
    Quantity & Functional form & Parameters \\
    \midrule
    \multicolumn{3}{l}{\textit{O-H stretch fundamentals}} \\
    \midrule
    Frequency
      & $\omega_{10} = a_0 + a_1 E + a_2 E^2$
      & $a_0 = 3760.2$ \\
      & & $a_1 = -3541.7$ \\
      & & $a_2 = -1.52677\times10^{5}$ \\[2pt]
    Dipole derivative
      & $\mu' = m_0 + m_1 E + m_2 E^2$
      & $m_0 = 0.1646$ \\
      & & $m_1 = 11.39$ \\
      & & $m_2 = 63.41$ \\[2pt]
    Position matrix element
      & $x_{10} = x_0 + x_1\,\omega_{10}$
      & $x_0 = 0.19285$ \\
      & & $x_1 = -1.7261\times10^{-5}$ \\[2pt]
    Momentum matrix element
      & $p_{10} = p_0 + p_1\,\omega_{10}$
      & $p_0 = 1.6466$ \\
      & & $p_1 = 5.7692\times10^{-4}$ \\[2pt]
    Intramolecular coupling
      & $\hbar\Omega^{\text{intra}} = [k_0 + k_1(E_i+E_j)]x_i x_j$
      & $k_0 = -1361.0$ \\
      & $\qquad\qquad + k_p\, p_i p_j$
      & $k_1 = 2.7165\times10^{4}$ \\
      & & $k_p = -1.887$ \\[2pt]
    Bond polarizability
      & $\alpha = \alpha_\perp \mathbf{I}
                 + (\alpha_\parallel-\alpha_\perp)\hat{u}\otimes\hat{u}$
      & $\alpha_\parallel = 0.185$ \\
      & 
      & $\alpha_\perp = 0.033$ \\[2pt]
    TDC dipole location
      & distance from O along O-H bond
      & $d = 0.67$~\AA \\
    \midrule
    \multicolumn{3}{l}{\textit{H-O-H bend overtones}} \\
    \midrule
    $0\!\rightarrow\!1$ frequency
      & $\omega^{10}_{\text{b}} = b_0 + b_1 E^{\text{b}}$
      & $b_0 = 1581.46$ \\
      & & $b_1 = 2938.51$ \\[2pt]
    $1\!\rightarrow\!2$ frequency
      & $\omega^{21}_{\text{b}} = c_0 + c_1 E^{\text{b}}$
      & $c_0 = 1551.32$ \\
      & & $c_1 = 3147.80$ \\[2pt]
    Fermi coupling
      & $\hbar\Omega^{\text{Fermi}} = W_{\text{F}}$
      & $W_{\text{F}} = 25~\&~0~\text{cm}^{-1}$ \\
    \bottomrule
  \end{tabular}
\end{table}
 

\paragraph{Constructing the vibrational exciton Hamiltonian.}
For a collection of $N$ chromophores in the selection
region, the vibrational exciton Hamiltonian at time $t$ is the
$N \times N$ matrix
\begin{equation}
  \kappa_{ij}(t) = \omega_i(t)\,\delta_{ij} + \Omega_{ij}(t)(1 - \delta_{ij}),
\end{equation}
where $\omega_i(t)$ is the local-mode fundamental frequency of
chromophore $i$, obtained from the electrostatic map, and
$\Omega_{ij}(t)$ is the vibrational coupling between chromophores $i$
and $j$. The chromophore basis comprises the O-H stretch fundamentals
together with the H-O-H bend overtones (one per water molecule), so
that the Fermi resonance between the stretch fundamental and the bend
overtone is treated explicitly (see below).\cite{Ni:2015aa,Kananenka:2018aa}

For two O-H stretches on the same water molecule, the
intramolecular coupling
is\cite{Auer:2008aa,Gruenbaum:2013aa}
\begin{equation}
  \hbar ~ \Omega_{ij}^{\text{intra}} = \left[k_0 + k_1(E_i + E_j)\right]\,x_i\,x_j + k_p\,p_i\,p_j,
\end{equation}
where $E_i$ is the projected electric field along O-H bond $i$, and $x_i$
and $p_i$ are the 0--1 coordinate and momentum matrix elements, taken
as functions of $\omega_i$.\cite{Gruenbaum:2013aa}  $k_0$, $k_1$, and
$k_p$ are map parameters derived from electronic structure
calculations.\cite{Gruenbaum:2013aa} For O-H stretch chromophores on different
molecules, the coupling is approximated by the transition dipole
coupling (TDC) mechanism,\cite{Auer:2008aa}
\begin{equation}
  \hbar \Omega_{ij}^{\text{inter}} = \frac{\mu_i'\,\mu_j'\,x_i\,x_j}{r_{ij}^3}\bigl[\hat{u}_i \cdot \hat{u}_j - 3(\hat{u}_i \cdot \hat{n}_{ij})(\hat{u}_j \cdot \hat{n}_{ij})\bigr],
\end{equation}
where $\mu_i'$ is the dipole derivative, $\hat{u}_i$ is the unit
vector along the OH bond, $\hat{n}_{ij}$ is the unit vector connecting
the transition dipole locations, and $r_{ij}$ is the distance between
them.

\paragraph{Fermi resonance and the bend overtone.}
The H-O-H bend overtone ($2\delta$) lies in the same spectral region as
the hydrogen-bonded O-H stretch ($\sim$3100--3300~\mbox{cm}$^{-1}$) and
mixes with it through a Fermi resonance, which has been shown to
contribute appreciably to the lineshape in this
range.\cite{Kananenka:2018aa,Ni:2015aa} We include this interaction by
augmenting the exciton basis with one bend-overtone chromophore for
each water molecule that falls within the
selection region. The diagonal frequency of the bend overtone is
obtained from the bend frequency map of Ni and
Skinner,\cite{Ni:2015aa}
\begin{equation}
  \omega^{2\delta}_i = \omega^{10}_{\text{b}}(E^{\text{b}}_i) + \omega^{21}_{\text{b}}(E^{\text{b}}_i),
\end{equation}
where $\omega^{10}_{\text{b}}$ and $\omega^{21}_{\text{b}}$ are the
$0\!\rightarrow\!1$ and $1\!\rightarrow\!2$ bend transition
frequencies, each a linear function of the bend electrostatic
collective coordinate $E^{\text{b}}_i$. Following Ni and
Skinner,\cite{Ni:2015aa} this coordinate is the sum of the components
of the electric field perpendicular to the two O-H bonds, projected
toward the H-O-H bisector and evaluated at the hydrogen atoms,
\begin{equation}
  E^{\text{b}}_i = \frac{\hat{e}_{\perp,1}\cdot\vec{E}(\mathrm{H}_1)}{r_{\mathrm{OH},1}}
                 + \frac{\hat{e}_{\perp,2}\cdot\vec{E}(\mathrm{H}_2)}{r_{\mathrm{OH},2}},
\end{equation}
where $\hat{e}_{\perp,k}$ is the in-plane unit vector perpendicular to
O-H bond $k$. The bend overtone is coupled to each of the two O-H
stretches on its parent molecule by a constant Fermi coupling
$\Omega^{\text{Fermi}} = W_{\text{F}}/\hbar$, with
$W_{\text{F}} = 25$~\mbox{cm}$^{-1}$ taken from Kananenka and Skinner.\cite{Kananenka:2018aa}  Again, following Ni and Skinner,\cite{Ni:2015aa}
intermolecular coupling between bend chromophores is neglected, as the
small bend transition dipole renders these couplings negligible
relative to the stretch couplings. The bend overtone carries no
independent transition dipole or polarizability in our implementation;
its spectroscopic intensity is borrowed entirely from the O-H stretch
fundamentals through the Fermi mixing encoded in the exciton
eigenvectors.

\paragraph{SFG spectra under the time-averaging approximation.}
At each MD configuration, the exciton Hamiltonian is diagonalized,
\begin{equation}
  \boldsymbol{\kappa} = \mathbf{V}\,\boldsymbol{\omega}\,\mathbf{V}^T,
\end{equation}
yielding eigenfrequencies $\omega_a$ and orthonormal eigenvectors with
components $V_{ia}$ giving the amplitude of local mode $i$ in
delocalized eigenstate $a$. The transition dipole and transition
polarizability of each eigenstate are obtained by projecting the
local-mode quantities onto the eigenvector,
\begin{equation}
  \mu^a_r = \sum_i V_{ia}\,\mu_{r,i}, \qquad
  \alpha^a_{pq} = \sum_i V_{ia}\,\alpha_{pq,i},
\end{equation}
where $\mu_{r,i}$ is the $r$-component of the transition dipole of
local mode $i$ and $\alpha_{pq,i}$ is the $pq$-component of its bond
polarizability tensor, with the Cartesian indices
$p,q,r \in \{x,y,z\}$ labeling laboratory-frame components. Each choice
of $(p,q,r)$ corresponds to a measurable polarization combination; for
example, $(p,q,r)=(y,y,z)$ contributes to the $ssp$ spectrum discussed
below.
 
In the time-averaging approximation, the dynamics within the
vibrational lifetime are assumed to be slow compared to the
inhomogeneous frequency spread, so each eigenstate contributes a
homogeneously broadened line centered at its eigenfrequency. The
resonant second-order susceptibility is then a sum of complex
Lorentzians,\cite{Auer:2007aa,Auer:2008aa}
\begin{equation}
  \chi^{(2)}_{pqr}(\omega) \propto
    \left\langle \sum_a
      \frac{\alpha^a_{pq}\,\mu^a_r}{\omega - \omega_a - i\Gamma}
    \right\rangle,
\end{equation}
where $\Gamma$ is the homogeneous half-width (a phenomenological
dephasing/lifetime parameter) and the angle brackets denote an average
over MD configurations. 
In the limit $\Gamma \to 0$ this reduces to the spectral density,
$\mathrm{Im}\,\chi^{(2)}_{pqr}(\omega) \propto \langle \sum_a
\alpha^a_{pq}\mu^a_r\,\delta(\omega - \omega_a)\rangle$, in which the
SFG intensity at frequency $\omega$ is the configuration-averaged sum
of eigenstate amplitudes resonant there.\cite{Auer:2008aa,Ni:2015aa}
A Bose--Einstein quantum correction
factor,
\begin{equation}
  Q(\omega) = \frac{\omega/k_BT}{1 - e^{-\omega/k_BT}}
\end{equation}
is applied to the computed susceptibility to account for detailed
balance.\cite{Kananenka:2018aa}
For the $ssp$ polarization combination (interface normal along $z$), the
relevant response is
\begin{equation}
  \chi^{(2)}_{ssp}(\omega) \propto \text{Im}\left[\frac{1}{2}\bigl(\chi^{(2)}_{xxz}(\omega) + \chi^{(2)}_{yyz}(\omega)\bigr)\right],
\end{equation}
where the two equivalent in-plane components are averaged to improve
statistics. The $sps$ and $ppp$ combinations are obtained
similarly.\cite{Kananenka:2018aa}
 
We note that the time-averaging approximation neglects motional
narrowing: because each configuration is treated as static, the
computed linewidths are governed by the inhomogeneous distribution of
eigenfrequencies together with the phenomenological width $\Gamma$,
and not by the averaging of frequency fluctuations over the vibrational
lifetime. For the O-H stretch region of water, where the lineshape is
predominantly inhomogeneously broadened, this is a reasonable
approximation,\cite{Auer:2008aa} and it is substantially less
expensive than explicit propagation of the exciton coherences, since
it requires only a single diagonalization of the exciton Hamiltonian
per configuration.

\subsection{O-H vibrational frequency distributions in additional ice bilayers}

\begin{figure}
    \centering
    \includegraphics[width=0.7\linewidth]{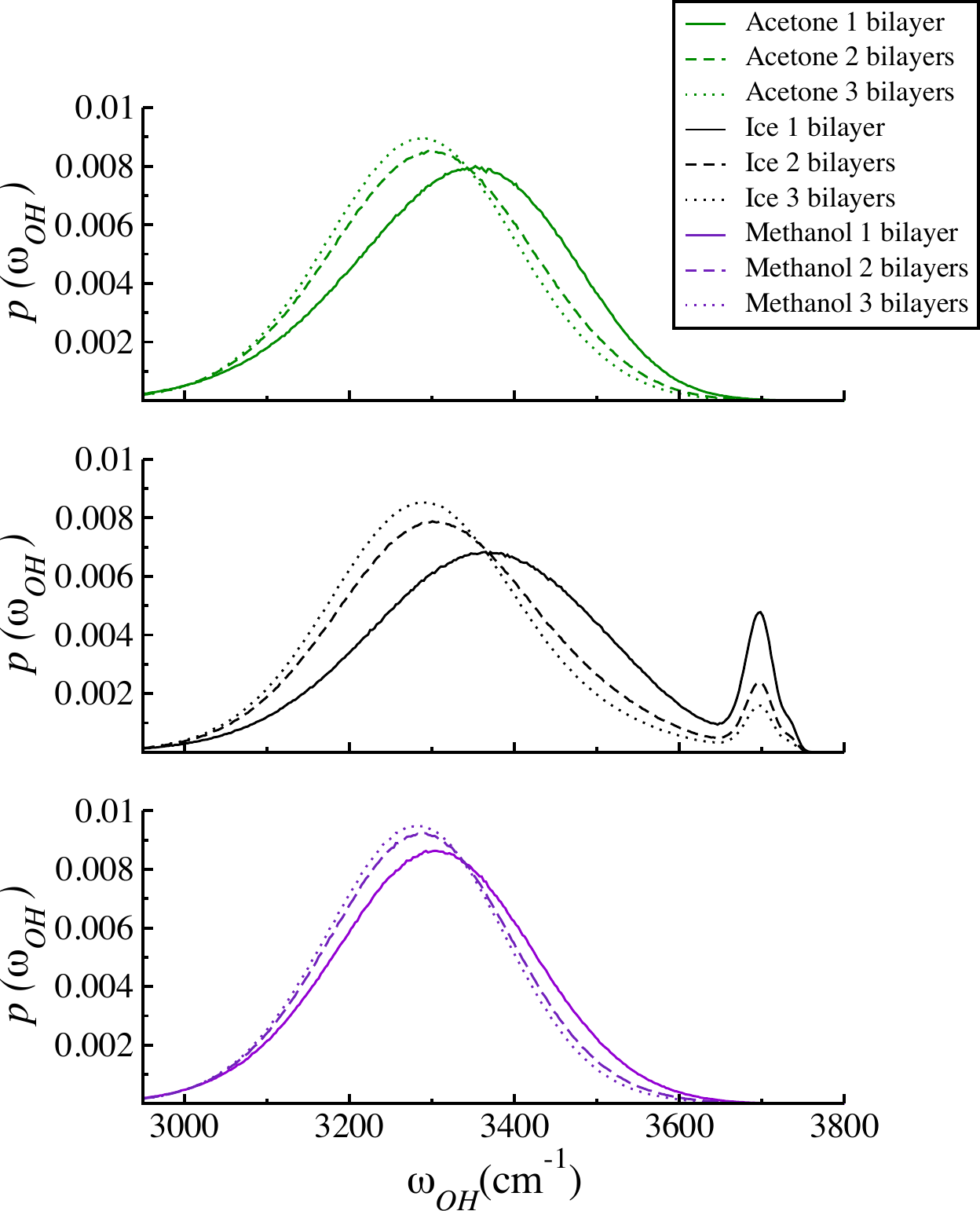}
    \caption{Frequency distributions of O-H stretching modes when including additional bilayers of ice at 223~K.}
    \label{fig:Bilayers}
\end{figure}

In the main paper, we provided probability distributions of O-H bond frequencies for water in the outermost ice bilayers at 223~K. In figure \ref{fig:Bilayers} we include additional bilayers in calculating these frequency distributions. We find that the inclusion of additional bilayers red-shifts the peak center for the acetone and ice samples, bringing them into alignment with the central frequency of the methanol sample. This suggests that the outermost water bilayer with a methanol adsorbate has more `bulk-like' electrostatic environments that drive O-H vibrations to lower frequency regions.

\subsection{Hydrogen Bonding}
In the main paper, we provided a table of hydrogen bonding statistics, \textit{i.e.}, probabilities of a water molecule being observed in a particular hydrogen bonding pattern. In table \ref{tab:FreeHBondCount}, we supply the average count of molecules sampled from 1000 configurations over the 1~ns data collection simulations.

\begin{table}
    \centering
    \caption{Average count of water molecules contributing to each hydrogen bonding class in the $67.35 \times 66.06 \mbox{~\AA}^2$ ice / adsorbate interfaces. For the 0~K and 223~K systems, this includes water molecules in the outermost bilayers on both sides of the slab. For the 265~K systems, this includes water in the outer half of $w_{\ce{H2O}}$. Molecules that under-donate ($n_D = 1$) are identified as DAA water molecules, while those that under-accept ($n_A=1$) are identified as DDA.}
\label{tab:FreeHBondCount}
\begin{tabular}{ l | l l | l l | l l }
                \toprule
 \multirow{2}{*}{Interface} & \multicolumn{2}{c|}{Hydrogen Bonding} & \multicolumn{2}{c|}{Donating} & \multicolumn{2}{c}{Accepting} \\ \cline{2-3} \cline{4-5} \cline{6-7} 
  & nHB $<$ 4 & nHB~=~4 & nD~=~1  & nD~=~2  & nA~=~1 & nA~=~2 \\
\midrule
\multicolumn{7}{l}{T = 0K, $N_{\ce{H2O}} = 1080$}\\ \midrule
 $\{0 0 0 1 \}$ & 540 & 540  & 270 & 810 & 270 & 810\\ \midrule
\multicolumn{7}{l}{T = 223~K, $\langle N_{\ce{H2O}}\rangle = 1080$}\\ \midrule
ice\textbar vapor & $380 \pm 3$  & $695 \pm 3$  & $200 \pm 2$ & $878 \pm 2$  & $212 \pm 2$ & $861 \pm 2$ \\
ice\textbar ace & $274 \pm 3$ & $806 \pm 2$ & $6.4 \pm 0.5$ & $1074.2 \pm 0.5$  & $268 \pm 3$ & $811 \pm 2$ \\
ice\textbar meoh & $101 \pm 3$ & $980 \pm 10$ & $5.7 \pm 0.3$ & $1080 \pm 10$  & $97 \pm 3$ & $990 \pm 10$ \\ \midrule
\multicolumn{7}{l}{T = 265~K, $\langle N_{\ce{H2O}}^\mathrm{ice}\rangle = 2300 \pm 200$, $\langle N_{\ce{H2O}}^\mathrm{ace}\rangle = 3400 \pm 200$,  $\langle N_{\ce{H2O}}^\mathrm{meoh}\rangle = 2900 \pm 400$} \\ \midrule
QLL & $550 \pm 20$ & $1700 \pm 200$ & $283 \pm 8$ & $2000 \pm 200$ & $350 \pm 10$ & $1900 \pm 200$ \\
QLL\textbar ace & $740 \pm 20$ & $2600 \pm 200$ & $202 \pm 9$  & $3200 \pm 200$  & $590 \pm 20$  & $2700 \pm 200$  \\
QLL\textbar meoh & $560 \pm 60$ & $2300\pm 400$ & $170 \pm 20$ & $2800 \pm 400$  & $440 \pm 40$ & $2400 \pm 400$ \\

\bottomrule
\end{tabular}
\end{table}

Note that a given water molecule can be undercoordinated in two ways: it can simultaneously donate and accept one hydrogen bond. For this reason, the counts of fully coordinated and undercoordinated waters can exceed the total number of water molecules in the sample. For the 0~K and 223~K systems, water in the sampling region comprised $\sim 1080$ molecules. However, due to the fluctuations in QLL widths, there was significantly more variance in molecular counts in the high temperature interfaces.

\bibliography{CiteKey}